%% file: main.tex
\newcommand{\isArxiv}

\documentclass[conference]{IEEEtran}
\usepackage{graphicx} 
\usepackage{amsmath,amssymb}

\ifCLASSINFOpdf
\else
\fi
\ifdefined\isArxiv
\usepackage{fancyhdr}
\fancypagestyle{firstpage}{%
  \rhead{Article under review for IEEE GLOBECOM ’22}
}
\fi

\begin{document}
%
\title{TreeStep: Tree Search for Vector Perturbation Precoding under per-Antenna Power Constraint}

\author{\IEEEauthorblockN{Abhishek Kumar Singh, \textit{Student Member IEEE}}
\IEEEauthorblockA{\textit{Princeton University}}
\and
\IEEEauthorblockN{Kyle Jamieson, \textit{Senior Member IEEE}}
\IEEEauthorblockA{\textit{Princeton University}}}


%


\maketitle
\ifdefined\isArxiv
\thispagestyle{firstpage}
\fi
\begin{abstract}
    Vector Perturbation Precoding (VPP) can speed up downlink data transmissions in Large and Massive Multi-User MIMO systems but is known to be NP-hard. While there are several algorithms in the literature for VPP under total power constraint, they are not applicable for VPP under per-antenna power constraint. This paper proposes a novel, parallel tree search algorithm for VPP under per-antenna power constraint, called \emph{\textbf{TreeStep}}, to find good quality solutions to the VPP problem with practical computational complexity. We show that our method can provide huge performance gain over simple linear precoding like Regularised Zero Forcing. We evaluate TreeStep for several large MIMO~($16\times16$ and $24\times24$) and massive MIMO~($16\times32$ and $24\times 48$) and demonstrate that TreeStep outperforms the popular polynomial-time VPP algorithm, the Fixed Complexity Sphere Encoder, by achieving the extremely low BER of $10^{-6}$ at a much lower SNR.
\end{abstract}

\IEEEpeerreviewmaketitle

\input{introduction}
\input{systemModel}
\input{relatedWork}

\input{design}
\input{evaluation}
\input{conclusion}

\bibliographystyle{IEEEtran}
\bibliography{biblio}

\end{document}

%% file: introduction.tex
\section{Introduction}
\label{sec:Intro}
Over the last few decades, we have seen a tremendous increase in the demand for data traffic, primarily dominated by video streaming. Online video streaming platforms like Netflix, Amazon Prime Video, YouTube, Twitch, etc., have seen their user count go up massively. We have also seen an increase in the capabilities of mobile devices, with current smartphones being able to support high-quality video and even VR services. As a result, modern wireless communication systems like LTE, 802.11ax, and 5G New Radio (NR) focus on providing high-speed downlink data to meet this growing user demand on mobile devices.

Current wireless systems are moving towards employing massive/large MIMO and MU-MIMO in the downlink to support this sustained user demand. There is also a drive toward reducing the overall end-to-end latency. As a result, coupled with an increase in the number of antennas on the AP, we also see a decrease in the timing budget for processing available to the AP. As a result, employing techniques that provide superior BER performance while remaining computationally cheap is the key to meeting the growing user demand.

While techniques like Dirty Paper Coding \cite{dpc} can achieve channel capacity, their computational complexity renders them impractical in modern wireless systems. Linear precoding techniques like Regularized Zero Forcing (RZF) are computationally cheap but offer inferior error performance. Vector Perturbation Precoding (VPP) \cite{vpPaper} is another technique that can provide a significant improvement over linear precoding. However, VPP is NP-Hard, and hence it is infeasible in most practical implementations. There are several approximate algorithms for VPP under total power constraint like the Fixed-Complexity Sphere Encoder (FSE)~\cite{vpFSE} and Degree-2 Sparse Vector Perturbation (D2VP)~\cite{sparseVP} that can achieve good BER performance with polynomial computational complexity. However, these methods rely on mathematical properties of the L2 norm and do not apply to VPP under per-antenna power constraint, which is an L-$\infty$ norm minimization problem. As pointed out in~\cite{perAntennaCons}, the per-antenna power constraint becomes relevant for modern MIMO systems with multiple RF chains, each with its own power amplifier.

This paper presents \textbf{\textit{TreeStep}}, a novel tree search algorithm to realize Vector Perturbation Precoding under per-antenna power constraint. TreeStep finds the VPP solution by performing a tree search that considers all possibilities for the first few levels but just a single possibility for each remaining level by performing local optimization operations. In contrast to FSE, which depends on QR decomposition to solve the VPP problem under total power constraint, TreeStep, described in Section~\ref{sec:design}, doesn't rely on the unitary invariance of L2 norm to perform an approximate tree search and hence can address an L-$\infty$ norm minimization. 

The rest of the paper is organized as follows. Section~\ref{sec:sysModel} describes the MIMO system model and VPP under per-antenna power constraint. Section~\ref{sec:relatedWork} talks about the existing work on VPP. Section~\ref{sec:design} presents the design of TreeStep. Section~\ref{sec:eval} contains extensive evaluation of TreeStep in various large/massive MIMO scenarios~($16\times16$, $24\times24$, $16\times32$, and $24\times48$). We demonstrate a significant performance gain over the popular FSE; for a $16\times16$ MIMO system using 4-, 16-, 64-QAM modulation, TreeStep can achieve the extremely low BER of $10^{-6}$ at $\approx 10$~dB lower SNR. We also perform extensive parameter tuning for TreeStep and demonstrate the impact of different parameters on performance and complexity. Section~\ref{sec:conclusion} concludes the paper and talks about possible directions for future work. 

%% file: systemModel.tex
\section{Vector Perturbation Precoding Model}
\label{sec:sysModel}
Consider downlink transmission in an MU-MIMO system with $N_{t}$ transmit antennas at the AP and one receive antenna at each user. The AP can transmit simultaneously to $N_{u}$ users. Let $\mathbf{x}$ be the transmit vector\footnote{For any vector $\mathbf{a}$, its $i^{th}$ element (scalar) is denoted by $a_{i}$. For any matrix $\mathbf{B}$, the scalar element in its $i^{th}$ row and $j^{th}$ column is denoted by $B_{ij}$}, with $x_{i}$ corresponding to the signal being transmitted from antenna $i$.  Let $y_{i}$ be the received signal at the $i^{th}$ user and $\mathbf{y} = (y_{1},y_{2}....y_{N_{u}})^T$ be the accumulated received vector. We assume a Rayleigh fading channel $\mathbf{H}$ and white Gaussian noise for each user with zero mean and standard deviation $\sigma$.
\begin{equation}
    \mathbf{y} = \mathbf{H}\mathbf{x} + \mathbf{n}
\end{equation}
where $\mathbf{n} = \mathcal{N}^{N_{u}\times 1}(0,\sigma^{2})$. Given a per-antenna power budget $P$, We define the signal to noise ratio, SNR($\rho$) as, 
\begin{equation}
    \rho = \dfrac{P}{E\left[||\mathbf{n}||^{2}\right]}
\end{equation}
In systems that use linear precoding, the transmitted signal is given by $\mathbf{x} = \mathbf{W}\mathbf{u}$ where $\mathbf{W}$ is the precoding matrix and $\mathbf{u}$ is the vector of symbols corresponding to each user. Each element of $\mathbf{u}$ is derived from a constellation $M$. For Zero Forcing Precoding~(ZF)~\cite{precodingSurvey}, 
\begin{equation}
    \mathbf{W} = \mathbf{H}^{\dag}(\mathbf{H}\mathbf{H}^{\dag})^{-1}
    \label{eq:precodMatZF}
\end{equation}
The transmit vector using ZF precoding and with a per-Antenna power budget of P is given by,
\begin{equation}
    \mathbf{x} = \sqrt{P} * \dfrac{\mathbf{H}^{\dag}(\mathbf{H}\mathbf{H}^{\dag})^{-1} \mathbf{u}}{||\mathbf{H}^{\dag}(\mathbf{H}\mathbf{H}^{\dag})^{-1} \mathbf{u}||_\infty}.
    \label{eq:transmitVec}
\end{equation}
We see that the effective SNR is inversely proportional to $||\mathbf{H}^{\dag}(\mathbf{H}\mathbf{H}^{\dag})^{-1} \mathbf{u}||_\infty$.
In Vector Perturbation Precoding \cite{vpPaper}, the transmit vector is given by $\mathbf{d} = \mathbf{u} + \tau \mathbf{v}$ where $\mathbf{u}$ is the vector of symbols corresponding to each user and $\mathbf{v} \in \mathcal{Z}^{N_{u} \times 1}$ is a vector of Gaussian integers. $\tau = 2(c_{max} + \Delta/2)$ where $c_{max}$ is the magnitude of the largest constellation symbol and $\Delta$ is the spacing between the constellation symbols \cite{vpPaper}. $\tau$ is a real number which is chosen to create non-overlapping shifted copies of the original constellation $M$. To achieve the best effective SNR, we choose $\mathbf{v} = \mathbf{v}^{\star}$ where,
\begin{equation}
    \mathbf{v}^{\star} = \text{arg} \min_{\mathbf{v}} ||\mathbf{H}^{\dag}(\mathbf{H}\mathbf{H}^{\dag})^{-1}(\mathbf{u} + \tau \mathbf{v})||_\infty.
\label{eq:vp}
\end{equation}
This is NP-Hard, and hence it is inefficient to solve it optimally on a wireless AP. For the remainder of this paper we will look at VPP as an optimization problem, where $\mathbf{W}$ corresponds to ZF precoding. However, our methods are applicable for any choice of $\mathbf{W}$. 
\begin{equation}
    \mathbf{v}^{\star} = \text{arg} \min_{\mathbf{v}} ||\mathbf{W}(\mathbf{u} + \tau \mathbf{v})||_\infty.
    \label{eq:mathvp}
\end{equation}

%% file: relatedWork.tex
\section{Related Work}
\label{sec:relatedWork}
Let us look at some existing work on efficiently solving the VPP problem. The sphere encoder described in \cite{vpPaper} is based on the Fincke and Pohst algorithm \cite{finPohst} and targets VPP under total power constraint. It involves expressing the matrix $\mathbf{W}$ in Equation \ref{eq:mathvp} as $\mathbf{W} = \mathbf{Q}\mathbf{R}$ by QR Decomposition, where $\mathbf{Q}$ is unitary and $\mathbf{R}$ is upper triangular. It then performs a tree search, assisted by the upper triangular structure \cite{finPohst}. The algorithm avoids exhaustively searching over the space of all integers by limiting the search to the points within a hyper-sphere of a suitably chosen radius. As the name implies, the algorithm is very similar to the Sphere Decoder algorithm used for the MIMO receiver. The Sphere Encoder provides optimal performance at a reduced complexity than the exhaustive search. Since both Sphere Encoder and Decoder are very similar algorithms, we can expect them to have similar behavior in terms of run time. In \cite{sphereDecoderComplexity}, it is shown that the expected complexity of the Sphere Decoder is exponential in the number of users, and hence techniques based on the Fincke and Pohst algorithm are feasible for problems of smaller sizes only \textit{i.e.}, a smaller number of users. The execution time of the sphere encoder algorithm varies across different instances of the VPP problem, which leads to several design challenges in practical systems. However, due to the lack of unitary invariance of the L-$\infty$ norm, the algorithm is not applicable. In~\cite{p-sphere}, authors present a modification of the Sphere decoder algorithm for p-norm minimization, which can be used for VPP under per antenna power constraint but requires exponential complexity (like the sphere encoder). Similarly, in~\cite{infinityNormApprox}, authors approximate L-$\infty$-norm minimization by minimization over L2 or higher norms. 

There are several algorithms in the literature that target VPP under total power constraint. A scheme for reducing the search space of the sphere encoder is discussed in \cite{modLoss}. It uses a heuristic of restricting the values of each perturbation to be drawn from a set of 4 possible values (two possible values for both real and imaginary parts) and hence reducing the search space of the sphere encoder. However, the search space is still exponential in size with respect to the number of users and the execution time of the algorithm varies across the instances of the same size. In \cite{decopVP}, the authors present an approximation to the VPP problem by minimizing the real and imaginary parts of the VPP cost function separately and hence reducing the search space of the sphere encoder. However, the search space is still exponential in size with respect to the number of users. 

We saw that tree-search algorithms based on the Fincke and Pohst algorithm \cite{finPohst} are inefficient for large MIMO systems, and hence there is a need for linear/polynomial-time algorithms for VPP. 
The Fixed Complexity Sphere Encoder (FSE) adapts the Sphere Encoder to have a lower and fixed complexity but leads to degradation in error performance \cite{vpFSE}. The Thresholded Sphere Search algorithm described in~\cite{vppThreshOpt} builds upon the sphere encoder algorithm~\cite{vpPaper} by adding additional stopping criteria in terms of an SNR dependent threshold heuristic. The performance and complexity of this algorithm depend inversely based on the choice of threshold. This algorithm also has variable execution time not only across instances of the same size but also for different SNRs. Degree-2 Sparse Vector Perturbation (D2VP), a low complexity algorithm for vector perturbation precoding, is presented in~\cite{sparseVP}. The algorithm reduces the complexity of finding the VPP solution by assuming that only two elements of the perturbation vector can be non-zero and then improves the solution via an iterative approach. This algorithm also has a polynomial-time complexity with respect to the number of users and does not do a QR decomposition-based tree search; however, it has a sequential execution and is not suitable for practical deployments which rely on parallel processing to support high data rates.

%% file: design.tex
\section{Design}
\label{sec:design}
We propose a new method for solving the VPP problem under per-antenna power constraint. The optimization problem that we have is 
\begin{equation}
    \mathbf{v}^{\star} = \text{arg} \min_{\mathbf{v}} ||\mathbf{W}(\mathbf{u} + \tau \mathbf{v})||_\infty,
     \label{eq:designProb}
\end{equation}
where $\mathbf{W}$ is the precoding matrix given by~(\ref{eq:precodMatZF}), $\mathbf{u}$ is the original transmit vector, and $\mathbf{v}$ is the integral perturbation vector. 
Since $\mathbf{v}$ takes only discrete values, gradient descent cannot be applied to find the optimal solution. However, the authors are motivated by the philosophy of gradient descent and develop an algorithm that improves the cost function of the given optimization problem after every iteration.

We can convert~(\ref{eq:designProb}) into an equivalent problem, where $\mathbf{u}$ and $\mathbf{v}$ are real valued, by reformulating the problem using the following transform for $\mathbf{W}$, $\mathbf{u}$, and $\mathbf{v}$:
\begin{equation}
    \mathbf{\hat{v}}^{\star} = \text{arg} \min_{\mathbf{\hat{v}}} ||\mathbf{\hat{W}}(\mathbf{\hat{u}} + \tau \mathbf{\hat{v}})||_\infty
     \label{eq:designProbReal}
\end{equation}
where,
\begin{equation}
\mathbf{\hat{W}}=
  \left[ {\begin{array}{cc}
   \Re(\mathbf{W}) +i\Im(\mathbf{W}) & -\Im(\mathbf{W})+ i\Re(\mathbf{W})\\
  \end{array} } \right],
\end{equation}
\begin{equation}
    \mathbf{\hat{u}}=
  \left[ {\begin{array}{c}
   \Re(\mathbf{u}) \\
   \Im(\mathbf{u}) \\
  \end{array} } \right],\text{~}\mathbf{\hat{v}}=
  \left[ {\begin{array}{c}
   \Re(\mathbf{v}) \\
   \Im(\mathbf{v}) \\
  \end{array} } \right],
  \label{eq:realTranVec}
\end{equation}
 with $\mathbf{\hat{u}}$ and $\mathbf{\hat{v}}$ all real. We re-construct the complex solution by inverting this transformation. 
\subsection{Local minimization along a single dimension}
\label{sec:localMin}
In this section, we will show that it is possible to derive an analytical expression for local minimization of the VPP objective function in a suitably defined "neighbourhood". 

Let $\Omega$ \textit{be the set of all possible perturbation vectors} $\mathbf{v}$. We recall that after the transformation described above, each element of $\mathbf{v}$ is a real integer. We see from~(\ref{eq:designProb}), if $\mathbf{W}$ is $n \times n$ and $\mathbf{v}$ is $n
\times 1$ then $\Omega = \mathbb{C}^{n}$. For the equivalent problem in~(\ref{eq:designProbReal}), $\mathbf{\hat{W}}$ is $n \times 2n$ and $\mathbf{v}$ is $2n\times 1$.
We denote the $i^{th}$ element of any vector $\mathbf{\hat{v}}$ as the \textbf{$i^{th}$ \textit{dimension} of} $\mathbf{\hat{v}}$. 
Let us consider a simpler version of the problem in~(\ref{eq:designProbReal}), where $\hat{\mathbf{W}}$ is replaced by an $1\times 2n$ row-vector $\mathbf{\hat{w}}^T$,
\begin{equation}
    \mathbf{\hat{v}}^{\star} = \text{arg} \min_{\mathbf{\hat{v}}} ||\mathbf{\hat{w}}^T(\mathbf{\hat{u}} + \tau \mathbf{\hat{v}})||_\infty,
\end{equation}
which is same as,
\begin{equation}
    \mathbf{\hat{v}}^{\star} = \text{arg} \min_{\mathbf{\hat{v}}} ||\mathbf{\hat{w}}^T(\mathbf{\hat{u}} + \tau \mathbf{\hat{v}})||^2,
    \label{eq:vpL2}
\end{equation}
If we try to locally minimize this objective function for VPP along a single dimension \textit{i.e.} optimize over the $i^{th}$ dimension of the perturbation vector, then as we will show next, it is possible to derive an analytical expression for the same and hence perform local minimization in a single computational step that requires $O(n)$ additions and multiplications.

Let $\hat{v}_{i} = x$ for some $i \in [1,n]$, be the dimension over which we are trying to optimize and elements corresponding to the other dimensions are treated as constants during this local minimization step. Let $\mathbf{q}$ be a vector consisting of all constant elements of $\mathbf{\hat{v}}$, such that, $q_j = \hat{v}_j$ for $j \neq i$ and $q_i = 0$. Let $a = \mathbf{\hat{w}}^H(\mathbf{\hat{u}}+\tau \mathbf{\hat{q}})$. Since $\hat{v}_{j} = 0$ for $j \neq i$ and $\hat{v}_{i} = x$, we can reformulate~(\ref{eq:vpL2}) as a quadratic function
\begin{equation}
    x^{\star}_i = \text{arg} \min_{x} ||a + \tau \hat{w}_{i}x||^{2},
\end{equation}
which takes the form
\begin{equation}
   x^{\star}_i  =  \text{arg} \min_{x} \alpha x^{2} + \beta x + \gamma,
\end{equation}
for some $\alpha,\beta$ and $\gamma$. We know that the solution is
\begin{equation}
    x^{\star}_i  = round(\dfrac{-\beta}{2 \alpha}),\text{~}f^{\star}_i = \alpha (x^{\star}_i)^{2} + \beta x^{\star}_i + \gamma,
\end{equation}
where $round(.)$ function rounds a given value to the nearest integer. The locally optimal solution after local minimization along the $i^{th}$ dimension is given by $x^{\star}_i$ and locally optimal value of the objective function as  is given by $f^{\star}_i$
We will use a greedy heuristic based on $f^{\star}_i$ later. Along with the real transformation described by~(\ref{eq:realTranVec}), this procedure is equivalent to optimizing the real part or the imaginary part of the perturbation corresponding to a single user while keeping the others constant. 
\begin{figure*}[h]
    \centering
    \includegraphics[width = \linewidth]{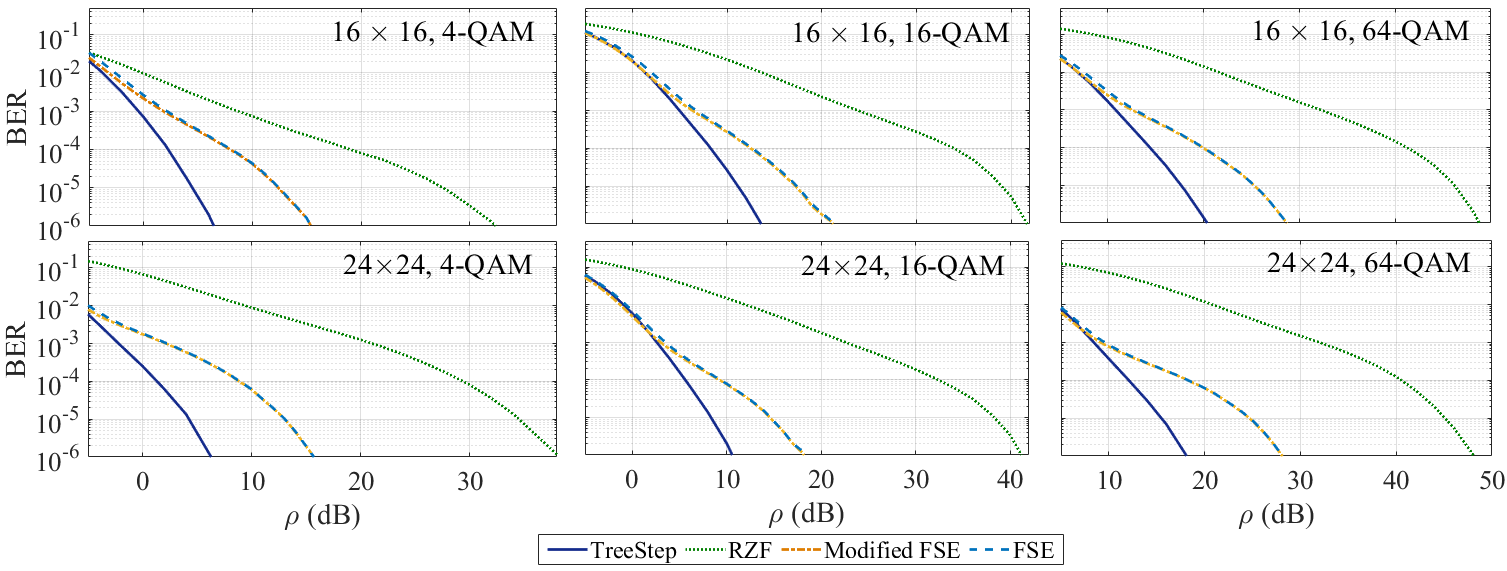}
    \caption{BER of the proposed TreeStep for Large MIMO systems: demonstrating performance gains over Fixed Complexity Sphere Encoder based VPP and Regularized Zero Forcing precoding. TreeStep parameters are set to $L=1,V=1,K=0$. We observe that both FSE and modified FSE perform similar, and TreeStep significantly improves their performance.}
    \label{fig:berLarge}
\end{figure*}
\subsection{TreeStep: Tree Search via local minimization steps}
\label{sec:TreeStepsub}
We present TreeStep, a fixed complexity tree search algorithm for VPP. Tree search algorithms for VPP like the Sphere Decoder~\cite{vpPaper} express the possible solutions in the form of the tree such that:\\ 
1. Children of root node correspond to all possible perturbations for dimension 1.\\
2. Children of a node at depth $i$ correspond to the all possible perturbations for dimension $i+1$.\\
3. The accumulation of the perturbation values corresponding to all nodes visited in the path from the root node to any leaf forms the perturbation vector.\\ 
The brute force traversal from root to each leaf node, which corresponds to checking every possible perturbation has exponential complexity in the number of users.

The key idea is to perform an approximate tree search by traversing the entire search tree up to a certain depth (full expansion) and for the remaining dimensions, optimizing along the dimensions which provides the best improvement to the objective function. (single expansion). 

Note that~(\ref{eq:designProbReal}) can be equivalently expressed as,
\begin{equation}
    \mathbf{\hat{v}}^{\star} = \text{arg} \min_{\mathbf{\hat{v}}} \max(g_1(\mathbf{\hat{v}}),g_2(\mathbf{\hat{v}}),...g_n(\mathbf{\hat{v}}))
\end{equation}

where $g_i(\mathbf{\hat{v}}) = ||\hat{w}_i^T(\mathbf{\hat{u}} + \tau \mathbf{\hat{v}})||^2$ and $\hat{w}_i^T$ is the $i^{th}$ row of the precoding matrix $\mathbf{\hat{W}}$. The key idea is to perform the local minimization procedure on each of these objective functions $g_i$ (which represents the power of $i^{th}$ antenna), and then select the local minimum which provides best improvement in the VPP objective function given by~(\ref{eq:designProbReal}). Let $x_{ij}^\star$ be the locally optimal solution obtained by minimizing $g_i$ along the dimension $j$. Let $\mathbf{\hat{v}^{ij}}$ be the perturbation vector with the $j^{th}$ element set equal to $x_{ij}^\star$, \textit{i.e.}, the perturbation vector that locally optimizes the power of the $i^{th}$ antenna along the dimension $j$. Let $P(\mathbf{\hat{v}}) = ||\mathbf{\hat{W}}(\mathbf{\hat{u}} + \tau \mathbf{\hat{v}})||_\infty$ be the power of the loudest antenna with perturbation $\mathbf{\hat{v}}$.

At each step, during single expansion, we fix one of the perturbation values. We compute locally optimal perturbation values $x_{ij}^\star$ for every antenna $i$, and every dimension $j$ that has not been fixed. We compute the corresponding values of $P(\mathbf{\hat{v}}^{ij})$, and select $x_{kl}^\star$ that leads to the smallest value of $P(\mathbf{\hat{v}})$. We fix the $l^{th}$ element of the perturbation vector to $x_{kl}^\star$ and repeat the procedure until all elements of the perturbation vector has been fixed. In the full expansion stage, we restrict each element of perturbation vector $\mathbf{\hat{v}}$ in~(\ref{eq:designProbReal}), such that $\hat{v}_{i} \in [-V,V]$. The perturbation values are most likely to be in $[-1,1]$, and setting $V = 1$ will suffice~\cite{perturbationDist}.

\subsubsection{\textbf{Stage 1}: Full Expansion for L Levels}
In this stage, we generate vectors corresponding to all possible perturbation values in $[-V,V]$ for dimensions 1 to $L$, \textit{i.e.} we do a full tree expansion for $L$ levels. This Step is similar to the FSE. At this stage, we generate $(2V+1)^L$ candidate solutions for the VPP problem, with dimensions 1 to $L$ of the perturbation vector assigned, in the form of a complete tree with degree 2$V$+1 and depth $L$. Each leaf node corresponds to a perturbation $\mathbf{q}$, which will serve as the starting vector for Step Descent.

\subsubsection{\textbf{Stage 2}: Single Expansion:}
In this stage, for every candidate solution generated in Stage~1, which have dimensions 1 to $L$ assigned, we use local minimization to find the perturbation values for the remaining dimensions. In terms of tree search, this is equivalent to a single expansion corresponding to every leaf node after Stage 1, \textit{i.e.} instead of traversing to all possible child nodes, we select just one child node for traversal. For every candidate solutions $\mathbf{q}$, generated by Stage~1, we use perform the following steps:\\
1. Start with $\mathbf{\hat{v}} = \mathbf{q}$.\\
2. Consider dimension $1$ to $L$ as assigned.\\
3. For every dimension $j$ (which has not yet been assigned a value) and antenna $i$, find the local minimum $x^{\star}_{ij}$ (as described in Section~\ref{sec:TreeStepsub}) and compute the corresponding value of $P(\mathbf{v}^{ij})$.\\
4. Let $P(\mathbf{\hat{v}}^{kl})$  be the smallest value  generated in Step~3. Assign $\hat{v}_{l} = x^{\star}_{kl}$ and mark dimension $l$ as assigned.\\
5. Repeat Steps 2 and 3 until all elements of $\mathbf{\hat{v}}$ have been assigned.\\
6. (Optional) Random Repetition:  Repeat the above steps $K$ times for the same problem starting with a random initial point such that in Step 1, elements of $\hat{v}_{i} = q_{i}$ for $i = 1$ to $L$  and other elements of $\mathbf{\hat{v}}$ are assigned random integral values in $[-B,B]$ where $B \leq V$. This step can be optionally executed if extra computational resources are available. In this paper, we set $B = V$.

We finally select the best solution among all the candidate solutions generated by the execution of Step Descent corresponding to each leaf node after Stage 1. 

\begin{figure*}[t!]
    \centering
    \includegraphics[width = \linewidth]{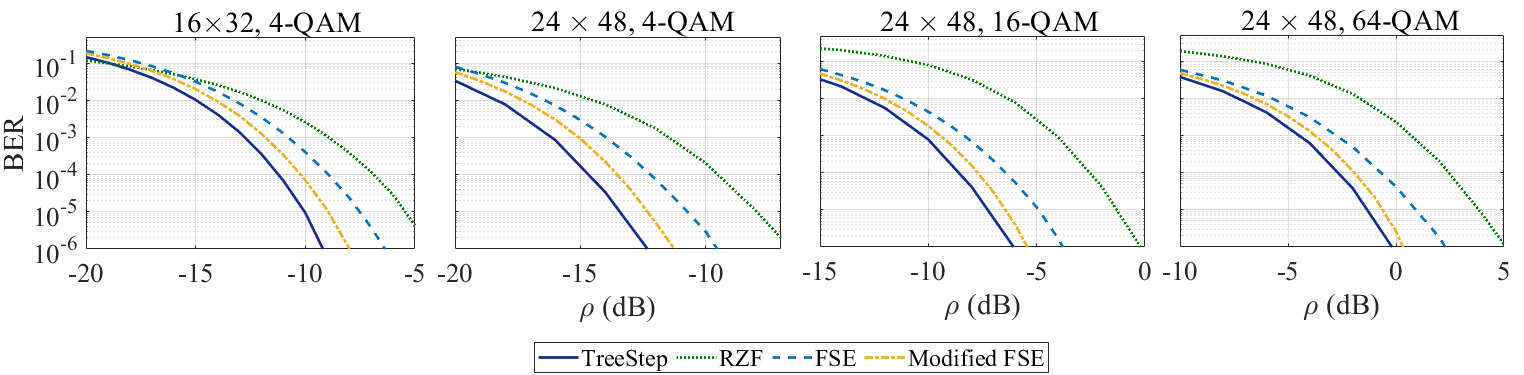}
    \caption{BER of the proposed TreeStep for Massive MIMO systems:  demonstrating performance gains over Fixed Complexity Sphere Encoder based VPP and Regularized Zero Forcing precoding. TreeStep parameters are set to $L=1,V=1,K=0$. We observe that modified FSE outperforms the conventional FSE for massive MIMO systems and TreeStep performs the best among of all evaluated methods.}
    \label{fig:berMassive}
\end{figure*}
\subsection{Computation Complexity and Parallelism}
In this section, we will look at the computational complexity of TreeStep. 
Let us say that for an $N_t\times N_r$ MIMO, we execute TreeStep with full expansion for $L$ levels searching all possible perturbations in the range $[-V,V]$, which generates $(2V+1)^L$ candidates. For each, of these candidates, we execute the single expansion for the remaining $2(N_t - L)$ depth. Note that the remaining depth is $2(N_t - L)$ and not $(N_t-L)$ as single expansion will used the transformation described by~(\ref{eq:realTranVec}). In the single expansion, for a node at depth $d$ performs $N_r(2N_t-d)$ minimization, each one of which requires O($N_t$) operations. So total number of operations required are 
\begin{equation}
    \sum_{d = 1}^{2(N_t-L)}d*(O(N_t)N_r) = O(N_rN_t(N_t-L)^2),
    \label{eq:complexity}
\end{equation}
hence the total complexity is $O((2V+1)^L N_r N_t (N_t-L)^2)$. If we include the optional random repetition option as well, then all these steps will be repeated $K$ times, and total complexity will be given  by $O((K+1)(2V+1)^L N_r N_t (N_t-L)^2)$. As noted before, we fix $V$, $L$ and $K$ to small values and hence the complexity of TreeStep is $O(N_rN_t^3)$, which is polynomial with respect to the MIMO size. Note that, that for large $N_t$, changing the parameters of TreeStep, will only modify, the complexity factor $C_f = (K+1)(2V+1)^L$ which characterises the depth and breadth of the full expansion stage, and random repetition. Note that, $C_f$ denotes the number of independent single expansions, each with the complexity of O($N_rN_t^3$), that need to be executed. Each of these $C_f$ single expansions are independent and can be executed in parallel or scheduled in any order over a pool of parallel computing resources. This allows for highly optimized GPU/FPGA implementations. In Section~\ref{sec:complexity}, we will look at the impact of changing various parameters on the BER performance with respect to the complexity factor $C_f$. 

%% file: evaluation.tex
\section{Evaluation}
\label{sec:eval}

In this section, we perform BER and spectral efficiency extensive evaluation of TreeStep. As noted before, the Fixed Complexity Sphere Encoder (FSE) was designed for the L2-norm minimization, and the technique doesn't extend to the L-$\infty$ norm due to the lack of unitary invariance property. For simulating the FSE, we perform the optimization operation assuming conventional VPP but use the per-antenna VPP criterion while transmitting symbols. We also evaluate FSE with a slight modification~(modified FSE), where we perform the optimization process assuming conventional VPP but use the per-antenna VPP criterion while selecting between various candidate solutions generated by FSE.

We simulate (using MATLAB) downlink MIMO transmission between a base station with $N_t$ transmit antennas and $N_r$ users with a single receive antenna each. We use a Rayleigh fading channel model and assume that the channel is accurately known at the base station. We simulate approximately $270\times 10^3$ MIMO instances (528 channel instances and 512 transmit vectors per channel). The scenario is equivalent to a $N_t \times N_r$ MIMO system, and its BER is computed as the average BER of the $N_r$ users. 
\subsection{Large MIMO Systems}
In this section, we look at the BER performance of TreeStep for large MIMO systems. Large MIMO systems have a comparable number of users and transmit antennas at the base station, and linear precoding techniques have terrible BER performance for such systems. In Fig.~\ref{fig:berLarge}, we simulate $16\times16$ MIMO and $24\times24$ MIMO with 4-, 16-, and 64-QAM modulations. We set TreeStep parameters to $L=1,V=1,K=0$. We observe that, as stated before, the performance of linear precoding (RZF) is bad, and it requires a high SNR to achieve satisfactory BER performance. We see that VPP based methods, Fixed Sphere Encoder (FSE) and TreeStep, provide a major improvement over the Regularised Zero-Forcing (RZF). Note that, for large MIMO systems, BER performance of the modified FSE is same as the FSE. We further see that our proposed scheme, TreeStep, provides significantly better BER performance than the Fixed Sphere Encoder. TreeStep can achieve the extremely low BER of $10^{-6}$ at a much lower SNR than the FSE. We can further improve the performance of TreeStep by using random repetition or increasing the depth of tree search; however, it will also increase the computational load. 
\subsection{Massive MIMO Systems}

Next, let us look at massive MIMO systems, which have a much higher number of transmit antennas~($N_t$) at the base station than the number of users~($N_u$). For such systems, the linear methods tend to have much better BER performance, especially when the ratio $\frac{N_t}{N_u}$ is high. Let us look at the BER performance of $16\times32$ and $24 \times 48$ massive MIMO systems in Fig.~\ref{fig:berMassive}. We set TreeStep parameters to $L=1,V=1,K=0$. We note that compared to large MIMO scenarios, the performance gap between RZF and VPP based methods (TreeStep and FSE) is much lower. We also see that, unlike large MIMO, modified FSE performs better than FSE. However, VPP based methods are still much better than ZF and our proposed scheme, TreeStep, provides the best BER performance out of the four tested precoders. 
\subsection{Performance vs Complexity: Parameter Tuning}
\label{sec:complexity}
\begin{figure}
    \centering
    \includegraphics[width=\linewidth]{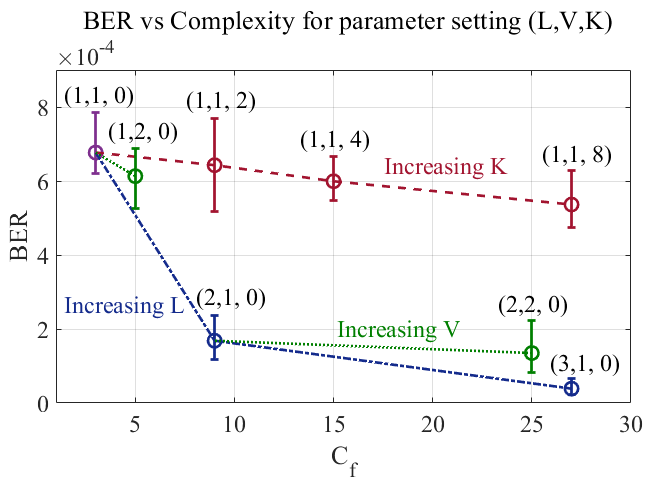}
    \caption{Variation of BER and Complexity~($8\times8$ MIMO, 4-QAM modulation, and 5~dB SNR) with variation in TreeStep parameters (L,V,K): L is the depth of full expansion, we search for optimal perturbation values is in [-V,V] during the full expansion stage, and K random repetitions are performed.}
    \label{fig:complexity}
\end{figure}

In this section, we try to understand the variation of BER with changes in the parameters of TreeStep. Recall that TreeStep has three parameters: the depth of full expansion ($L$), the width of full expansion ($V$), and the number of random repetitions~($K$) of the single expansion stage. In Fig.~\ref{fig:complexity}, we look at the BER of $8\times8$ MIMO with 4-QAM modulation, at $\rho = 5$~dB. The parameter setting of different of TreeStep is represented by the 3-tuple ($L,V,K$). Recall that $C_f = (K+1)(2V+1)^L$ denotes the number of independent single-expansions, each with the complexity of O($N_rN_t^3$), that need to be executed. We see that increasing the depth~($L$) of full expansion leads to significant performance gains; however, it causes a geometric increase in complexity. Increasing the width~($V$) has very little impact on the performance, as it is highly likely that the optimal perturbation values are contained in $[-1,1]$ and $V = 1$ is sufficient. 
Performing random repetitions of the single expansion stage can provide incremental gain in performance. However, the performance gains appear to be small compared to the increase in complexity, and increasing $L$ seems to provide much higher performance gains at a similar complexity cost. Therefore, random repetitions can be used when surplus processing resources are not enough for increasing $L$; otherwise, increasing $L$ seems to be a better strategy.

%% file: conclusion.tex
\section{Conclusion and Future work}
\label{sec:conclusion}
In this paper, we present TreeStep, a new tree search algorithm for Vector Perturbation Precoding~(VPP) under per-antenna power constraint. Unlike the existing state-of-the-art techniques, which target VPP under total power constraint, TreeStep decouples the search procedure from QR decomposition and unitary invariance of L2 norm. We show that this allows TreeStep to search for better solutions under per-antenna power constraint, and hence provide a much better BER performance.

While VPP can provide a huge BER improvement, its implementation in commercial systems remains infeasible due to its extremely high computational complexity. Approximation algorithms that embody parallelism, like TreeStep, can effectively solve the VPP problem and play a crucial role in making VPP suitable for commercial systems.  

The FSE implementations will usually employ certain heuristics to reorder the channel matrix before executing the algorithm to decide which dimensions receive the benefits of full expansion. Since TreeStep and FSE do not differ in the full expansion step, the same heuristics can be applied to TreeStep as well. We believe TreeStep can reap the same benefits from reordering as the FSE and then perform a superior neighborhood search, leading to an overall improvement in BER performance. We want to explore these performance enhancements of TreeStep with these heuristics.